\documentstyle[12pt,epsfig]{article}                                                                                       
                                                                                       
\newlength{\dinwidth}                                                                                       
\newlength{\dinmargin}                                                                                       
\def\lapproxeq{\lower .7ex\hbox{$\;\stackrel{\textstyle                                                                                       
<}{\sim}\;$}}                                                                                       
\def\gapproxeq{\lower .7ex\hbox{$\;\stackrel{\textstyle                                                                                       
>}{\sim}\;$}}                                                                                       
\def\be{\begin{equation}}                                                                                       
\def\ee{\end{equation}}                                                                                       
\def\bea{\begin{eqnarray}}                                                                                       
\def\eea{\end{eqnarray}}

\makeatletter                                                              
\def\fmslash{\@ifnextchar[{\fmsl@sh}{\fmsl@sh[0mu]}}                                                              
\def\fmsl@sh[#1]#2{%
\mathchoice                                                              
{\@fmsl@sh\displaystyle{#1}{#2}}%
{\@fmsl@sh\textstyle{#1}{#2}}%
{\@fmsl@sh\scriptstyle{#1}{#2}}%
{\@fmsl@sh\scriptscriptstyle{#1}{#2}}}                                                              
\def\@fmsl@sh#1#2#3{\m@th\ooalign{$\hfil#1\mkern#2/\hfil$\crcr$#1                                                              
#3$}}                                                              
\makeatother            
                                                                                       
\begin{document}                                                                                       
\titlepage                                                                                       
\begin{flushright}                                                                                       
DTP/99/18 \\                                                                                      
February 1999 \\                                                                                       
\end{flushright}                                                                                       
                                                                                       
\vspace*{2cm}                                                                                       
                                                                                       
\begin{center}                                                                                       
{\Large \bf Off-diagonal distributions fixed by diagonal} \\                                                           
                                                           
\vspace*{0.3cm}                                                           
{\Large \bf partons at small $x$ and $\xi$}                                                                                       
                                                                                       
\vspace*{1cm}                                                                                       
A.G.~Shuvaev$^{a}$, K.J.~Golec-Biernat$^{b,c}$, A.D.~Martin$^b$ and 
M.G.~Ryskin$^{a,b}$ \\                                                                                       
                                                                                      
\vspace*{0.5cm}                                                                                      
$^a$ Petersburg Nuclear Physics Institute, Gatchina, St. Petersburg 188350, Russia \\ 
$^b$ Department of Physics, University of Durham, Durham, DH1 3LE \\                                                                                      
$^c$ H.\ Niewodniczanski Institute of Nuclear Physics, 31-342 Krakow, Poland \\
\end{center}                                                                                       
                                                                                       
\vspace*{2cm}                                                                                       
                                                                                       
\begin{abstract}                                                                     
We show that the off-diagonal (or skewed) parton distributions are completely 
determined at small $x$ and $\xi$ by the (conventional) diagonal partons.  We present 
predictions which can be used to estimate the off-diagonal distributions at small $x$ 
and $\xi$ at any scale.
\end{abstract}                                                                                      
                                                                              
\newpage                                                                                      
\section{Introduction}                                          

Precision data are becoming available for hard scattering processes whose description 
requires knowledge of off-diagonal (or so-called \lq\lq skewed\rq\rq) parton 
distributions.  Particularly relevant processes are the diffractive production of vector 
mesons and of high $E_T$ jets in high energy electron-proton collisions.

We shall use the \lq\lq off-forward\rq\rq~distributions
$$
H (x, \xi) \; \equiv \; H (x, \xi, t, \mu^2)
$$
with support $-1 \le x \le 1$ introduced by Ji \cite{J1,J2,J3}, with the minor difference 
that the gluon $H_g = x H_g^{\rm Ji}$.  They depend on the momentum fractions 
$x_{1,2} = x \pm \xi$ carried by the emitted and 
absorbed partons at each scale $\mu^2$ and on the momentum transfer variable $t = 
(p - p^\prime)^2$, see Fig.~1.  The values of $t$ and $\xi = (x_1 - x_2)/2$ do not 
change as we evolve the parton distributions up in the scale $\mu^2$.  That is $t$ and 
$\xi$ lie outside the evolution.  
In the limit $\xi \rightarrow 0$ the distributions reduce to the conventional diagonal 
distributions
\bea
\label{eq:aa}
H_q (x, 0) & = & \left \{ \begin{array}{c} q (x) \quad\quad \;\; {\rm for} \quad x > 0 \\
- \bar{q} (- x) \quad {\rm for} \quad x < 0 \end{array} \right . \nonumber \\
& & \\
H_g (x, 0) & = & x \: g (x). \nonumber
\eea
Detailed reviews of off-diagonal distributions can be found, for example, in 
refs.~\cite{J3,R1,GM}.

It is usual to anticipate that the $\xi$ dependence of $H$ is controlled by the 
non-perturbative starting (input) distribution at some low scale $\mu^2 = Q_0^2$.  
However here we wish to explore the possibility that, in the small $x, \xi \ll 1$ region, 
the \lq\lq skewed\rq\rq~off-diagonal behaviour comes mainly from the evolution.  
Indeed we expect this to be the case.  At each step of the evolution the momentum 
fraction carried by parton $i$ ($i = 1,2$) decreases.  So when the evolution length is 
sufficiently large (i.e.~$\ln (Q^2/Q_0^2) \gg 1$), the important values of $x \sim x_0$ 
of the input $(\mu^2 = Q_0^2$), which control the behaviour in the $x \sim \xi$ 
domain at the high scale $(\mu^2 = Q^2)$, will satisfy $x_0 \gg \xi$.  Clearly we can 
neglect the $\xi$ dependence in the $x_0$ region and start evolving from purely 
diagonal partons with $x_1 = x_2$.

Here we demonstrate how, in the phenomenologically important small $\xi$ region 
(for $t \rightarrow 0$), the off-diagonal distributions are determined unambiguously in 
terms of the small $x$ behaviour of the (conventional) diagonal partons which is 
known from experiment.  We therefore have the attractive possibility to include data 
described by off-diagonal distributions in a global parton analysis to better constrain 
the small $x$ behaviour of the diagonal distributions. \\

\section{Off-diagonal distributions in terms of conformal moments}

In terms of the Operator Product Expansion (OPE) the evolution of the off-diagonal 
distributions may be viewed as the renormalisation of the matrix elements $O_N = 
\langle p^\prime | \hat{O}_N | p \rangle$ of the conformal (Ohrndorf \cite{O}) 
operators, where $p$ and $p^\prime$ are the momenta of the incoming and outgoing 
protons.  For the quark, the leading twist operator $\hat{O}_N$ is of the form
\be
\label{eq:a1}
\hat{O}_N^q \; = \; \sum_{k = 0}^N \: \left ( \begin{array}{c} N \\ k \end{array} 
\right ) \: \left ( \begin{array}{c} N + 2 \\ k + 1 \end{array} \right ) \: 
\partial_L^k \: \partial_R^{N - k}
\ee
where the derivatives $\partial_L$ and $\partial_R$ act on the left and right quarks in 
Fig.~1.  As a consequence the quark matrix element takes the form
\be
\label{eq:a2}
O_N^q \; = \; \int_{-1}^1 \: dx \: R_N^q (x_1, x_2) \: H_q (x, \xi)
\ee
with $x_{1,2} = x \pm \xi$, where the polynomials \cite{BFKL}
\be
\label{eq:a3}
R_N^q \; = \; \sum_{k = 0}^N \: \left ( \begin{array}{c} N \\ k \end{array} \right ) \: 
\left ( \begin{array}{c} N + 2 \\ k + 1 \end{array} \right ) \: x_1^k \: x_2^{N - 
k}.
\ee
In other words the polynomials $R_N (x_1, x_2)$ are the basis which specifies the 
conformal moments $O_N$.  In the diagonal limit, with $x_1 = x_2$, (\ref{eq:a2}) 
reduces to the well-known moments
\be
\label{eq:a4}
M_N^q \; = \; \int_0^1 \: x^N \: q (x) \: dx.
\ee
Unlike the common $x^N$ basis in the diagonal case, the gluon and quark polynomial 
bases differ from each other.  For the gluon we have
\be
\label{eq:a5}
R_N^g \; = \; \sum_{k = 0}^N \: \left ( \begin{array}{c} N \\ k \end{array} \right ) \: 
\left ( \begin{array}{c} N + 4 \\ k + 2 \end{array} \right ) \: x_1^k \: x_2^{N - 
k},
\ee
to be compared with the quark polynomials of (\ref{eq:a3}).

Recall that the off-diagonal distributions are symmetric in $\xi$ \cite{J2,J3}
\be
\label{eq:a6}
H_i (x, \xi) \; = \; H_i (x, - \xi)
\ee
with $i = q$ or $g$.  This is just the left-right or $x_1 \leftrightarrow x_2$ symmetry 
of Fig.~1.  In terms of the $x$ variable the symmetry relations are
\bea
\label{eq:a7}
H_q^s (x, \xi) & = & - H_q^s (-x, \xi), \nonumber \\
& & \nonumber \\
H_q^{ns} (x, \xi) & = & H_q^{ns} (-x, \xi), \\
& & \nonumber \\
H_g (x, \xi) & = & H_g (-x, \xi) \nonumber
\eea
for the quark singlet, non-singlet and gluon respectively.

The conformal moments $O_N$ have the advantage that they are not mixed, at least at 
LO, during the off-diagonal evolution, but simply get multiplicatively 
renormalized\footnote{For simplicity we take the coupling $\alpha_S$ to be fixed.  
The generalisation to running $\alpha_S$ is straightforward.}
\be
\label{eq:a8}
O_N (Q^2) \; = \; O_N (Q_0^2) \: \left ( \frac{Q^2}{Q_0^2} \right )^{\gamma_N}
\ee
with the same anomalous dimension $\gamma_N$ as in the diagonal case.  The 
problem of how to restore the analytic off-diagonal distribution $H (x, \xi)$ from 
knowledge of its conformal moments $O_N (\xi)$ of (\ref{eq:a2}) has been solved 
recently by Shuvaev \cite{SHUV}.  The prescription is as follows.  We first calculate 
an auxiliary function
\be
\label{eq:a9}
f (x^\prime, \xi; t) \; \equiv \; f (x^\prime) \; = \; \int \: \frac{dN}{2 \pi i} \: 
(x^\prime)^{-N} \: O_N (\xi)/R_N (1,1)
\ee
using a simple Mellin transform, where for simplicity of presentation we shall omit 
the arguments $\xi, t$ and $\mu^2$ of $f$.  Next we perform the convolution
\be
\label{eq:a10}
H (x, \xi) \; = \; \int_{-1}^1 \: dx^\prime \: {\cal K} (x, \xi; x^\prime) \: f (x^\prime),
\ee
where, for quarks, the kernel is given by
\be
\label{eq:a11}
{\cal K}_q (x, \xi; x^\prime) \; = \; - \frac{1}{\pi | x^\prime |} \: {\rm Im} \: 
\int_0^1 \: ds (1 - y(s) x^\prime)^{-3/2}
\ee
with
\be
\label{eq:a12}
y (s) \; = \; \frac{4s (1 - s)}{x + \xi (1 - 2s)}.
\ee

To gain insight into the Shuvaev prescription we repeat that, from a theoretical OPE 
point of 
view, it is best to analyse experimental data for processes described by off-diagonal 
distributions in terms of the conformal moments $O_N$ of (\ref{eq:a2}) which 
diagonalize the (LO) evolution.  However, phenomenologically it is more convenient 
to work in terms of the off-diagonal parton distributions themselves.  The Shuvaev 
transform (\ref{eq:a9}) and (\ref{eq:a10}) performs the necessary inverse of 
(\ref{eq:a2}) at any fixed $\xi, t$ and $\mu^2$; that is it enables $H (x, \xi)$ to be 
constructed from $O_N (\xi)$.  So far this is just a mathematical procedure.  The 
crucial physical step is to relate the auxiliary function $f (x^\prime)$ directly to the 
diagonal partons.  It is easy to show, for $\xi \ll 1$, that $f (x^\prime)$ in fact 
reduces to a diagonal parton distribution.  Indeed the conformal moments may be 
expressed in the form
\be
\label{eq:a13}
O_N (\xi) \; = \; \sum^{[(N + 1)/2]}_{k = 0} \: O_{Nk} \: \xi^{2k},
\ee
which embodies the \lq\lq polynomial condition\rq\rq~that the power of $\xi$ should 
be at most of the order of $N + 1$.  For $\xi \ll 1$ we have
\be
\label{eq:a14}
O_N (\xi) \; \simeq \; O_{N0} \; = \; O_N (0).
\ee
Now, up to the trivial normalization factor $R_N (1,1)$, the diagonal moment $O_N 
(0)$ is equal to the $x^N$ moment of the diagonal parton distribution.  So for $\xi \ll 
1$ we can put $f_q (x^\prime) = q (x^\prime)$ in (\ref{eq:a10}), and then use 
(\ref{eq:a11}) to determine the off-diagonal distribution $H_q (x, \xi)$ in terms of the 
conventional quark distribution.  In this limit the kernel ${\cal K}$ just becomes a 
non-trivial representation of the delta function $\delta (x - x^\prime)$.

Since (\ref{eq:a11}) is a principal value integration, the apparent singularity at $y (s) 
x^\prime = 1$ is not a problem.  However, for computation purposes, it is convenient 
to first weaken this singularity in the $s$ integration by integrating by parts.  Then 
(\ref{eq:a10}) and (\ref{eq:a11}) become
\be
\label{eq:a15}
H_q (x, \xi) \; = \; \int_{-1}^1 \: dx^\prime \left [ \frac{2}{\pi} \: {\rm Im} \: \int_0^1 
\: \frac{ds}{y (s) \: \sqrt{1 - y(s) x^\prime}} \right ] \: \frac{d}{dx^\prime} \left ( 
\frac{q (x^\prime)}{| x^\prime |} \right ).
\ee
Here we have used the properties that $q (x^\prime) \rightarrow 0$ as $x^\prime 
\rightarrow 1$ and that
\be
\label{eq:a16}
q^s (x^\prime) \; = \; -q^s (- x^\prime), \quad\quad q^{ns} (x^\prime) \; = \; q^{ns} (- 
x^\prime),
\ee
see (\ref{eq:a7}).  Note that, for small $\xi$, we can identify the auxiliary function $f 
(x^\prime)$ of (\ref{eq:a9}) with the diagonal partons at any scale, as the same 
anomalous dimensions $\gamma_N$ control both the diagonal and off-diagonal 
evolution.

So far we have neglected the $t$ dependence and set $t = 0$.  However from the sum 
rule \cite{J3} we know that the $t$ dependence of the first conformal moment is given 
by the proton form factor $G (t)$,
\be
\label{eq:a17}
O_{N = 0} (t) \; = \; \langle p^\prime | \hat{O}_0 | p \rangle \; \propto \; G (t).
\ee
In fact it is natural to assume that all the moments are proportional to $G (t)$
\be
\label{eq:a18}
O_N (t) \; = \; O_N (t = 0) \: G (t)
\ee
and simply multiply (\ref{eq:a10}) by $G (t)$ to restore the $t$ dependence of the 
distributions.  Another argument in favour of such a factorization is the form of the 
Mellin integration (\ref{eq:a9}) where, for small $x$, the saddle point is located near 
the singularity at $N = 0$ which comes from the behaviour of the singlet anomalous 
dimension, $\gamma_N \propto 1/N$.  Thus the dominant contribution comes from 
$O_{N = 0}$ which is indeed proportional to $G (t)$, and due to the polynomial 
condition (\ref{eq:a13}) does not depend on $\xi$ at all \cite{J3}.

The formula for the gluon is a little different to that for the quarks.  The reason is that 
in the off-diagonal case the functions $H_q$ and $H_g = xH_g^{\rm Ji}$ form the 
singlet multiplet which is multiplicatively renormalized.  The additional $x$ in the 
gluon reveals itself as an extra factor of $x + \xi (1 - 2s)$ in the kernel.  Thus for the 
gluon, in place of (\ref{eq:a15}), we have
\be
\label{eq:a19}
H_g (x, \xi; t) \; = \; x H_g^{\rm Ji} \; = \; \int_{-1}^1 \: dx^\prime \left [ \frac{2}{\pi} 
\: {\rm Im} \: \int_0^1 \: \frac{ds (x + \xi (1 - 2s))}{y (s) \: \sqrt{1 - y (s) x^\prime}} 
\right ] \: \frac{d}{dx^\prime} \left ( \frac{g (x^\prime)}{| x^\prime |} \right ) \: G (t).
\ee

\newpage
\section{Predictions of the off-diagonal distributions for small $x$ and $\xi$}

We see that (\ref{eq:a15}) and (\ref{eq:a19}) completely determine the behaviour of 
the off-diagonal distributions in the small $x, \xi$ domain in terms of the diagonal 
distributions.  In fact by making the physically reasonable small $x$ assumption that 
the diagonal partons are given by
\be
\label{eq:a20}
x q (x) \; = \; N_q \: x^{- \lambda_q}, \quad\quad xg (x) \; = \; N_g \: x^{- \lambda_g}
\ee
we can perform the $x^\prime$ integration analytically\footnote{We use the 
substitution $z = 1/x^{\prime} y(s)$ and note that $$ \int_0^1 \: dz \: z^{\lambda + 
\frac{3}{2}} \: (1 - z)^{- \frac{1}{2}} \; = \; \frac{\Gamma \left ( \lambda + 
\frac{5}{2} \right ) \: \Gamma \left ( \frac{1}{2} \right )}{\Gamma (\lambda + 
3)}.$$}.  We obtain
\be
\label{eq:a21}
H_i (x, \xi; t) \; = \; N_i \: \frac{\Gamma \left ( \lambda + \frac{5}{2} \right 
)}{\Gamma (\lambda + 2)} \: \frac{2}{\sqrt{\pi}} \: \int_0^1 \: ds \left [ x 
+ \xi (1 - 2s) \right ]^p \: \left [ \frac{4s (1 - s)}{x + \xi (1 - 2s)} \right ]^{\lambda_i + 
1} \: G (t)
\ee
with $i = q$ or $g$, and where $p = 0$ and 1 for quarks and the gluon respectively.

At first sight it appears that for singlet quarks (where $\lambda_q > 0$ and $p = 0$) 
we face a strong singularity in integral (\ref{eq:a21}) when the term $D \equiv x + \xi 
(1 - 2s) \rightarrow 0$ in the denominator.  Fortunately the singlet quark distribution 
is antisymmetric in $x$.  To obtain the imaginary part of the integral (\ref{eq:a15}) 
we must choose $x^\prime > 0$ for $D > 0$ and $x^\prime < 0$ for $D < 0$.  
Therefore we must treat (\ref{eq:a21}) as a principal value integral and take the 
difference between the $D \rightarrow 0+$ and $D \rightarrow 0-$ limits.  Thus the 
main singularity is cancelled and (\ref{eq:a21}) becomes integrable for any 
$\lambda_q < 1$.

Note that the dominant contribution to the $x^\prime$ integrations of (\ref{eq:a15}) 
and (\ref{eq:a19}) comes from the region of small $x^\prime \sim x, \xi$.  Indeed 
with the input given by (\ref{eq:a20}), the integral for the quark distribution 
has a strong singularity at small 
$x^\prime$
\be
\label{eq:b21}
I_q \; \sim \; \int \: dx^\prime (x^\prime)^{- \lambda_q - 3} \: {\rm Im} \left ( 
\frac{1}{y (s) \sqrt{1 - y (s) x^\prime}} \right ).
\ee
However when we take the imaginary part, the $x^\prime$ integration is cut-off by the 
theta function $\theta (x^\prime - 1/y (s))$ at
\be
\label{eq:c21}
x^\prime \; = \; 1/y (s) \; \sim \; x + \xi (1 - 2s).
\ee
So we obtain the small $\xi$ behaviour $I_q \sim \xi^{- \lambda_q - 1}$, and the 
distribution (\ref{eq:a15}) has the form
\be
\label{eq:d21}
H_q (x, \xi) \; = \; \xi^{- \lambda_q - 1} \: F_q (x/\xi).
\ee
Similarly it follows that $H_g = \xi^{- \lambda_g} F_g (x/\xi)$.

The predictions for the off-diagonal distributions are shown in Fig.~2.  In diagrams 
(a)--(c) we show the ratio $R$ to the diagonal distribution in the form
\be
\label{eq:a22}
R \; = \; \frac{H (x, \xi)}{H (x + \xi, 0)},
\ee
and so the only free parameter is $\lambda$, the exponent which fixes the $x^{- 
\lambda}$ behaviour of the input diagonal partons, as in (\ref{eq:a20}).  Notice that 
on account of (\ref{eq:d21}) the ratios $R$ at small $x$ and $\xi$ are a function of 
only the ratio of the variables $x/\xi$.

The ratios $R$ of (\ref{eq:a22}) are the relevant ratios.  For example, high energy 
diffractive $q\bar{q}$ electroproduction is described by two gluon exchange with
\be
\label{eq:a23}
x_1 \; \simeq \; (Q^2 + M_{q\bar{q}}^2) \: \big{/} \: W^2 \gg x_2,
\ee 
where $W$ is the centre-of-mass energy of the proton and the photon of virtuality 
$Q^2$.  A common approximation is to describe the process in terms of the diagonal 
gluon $x_1 g (x_1)$, sampled at $x_1 = x + \xi$.  In this case the inclusion of 
off-diagonal effects will enhance the cross section by a factor of $R_g^2$, where 
$R_g$ is evaluated at $x/\xi = 1$, see Fig.~2(b) or 2(d).

For $x \gg \xi$ we see that the off-diagonal to diagonal ratios, $R$, tend to unity, as 
expected.  Moreover, due to the $x \rightarrow -x$ antisymmetry property 
(\ref{eq:a7}), we see that the quark singlet vanishes as $x \rightarrow 0$.  Also for a 
flat input gluon, $xg (x) \rightarrow$ constant as $x \rightarrow 0$ (that is $\lambda_g 
= 0$), we see that $R_g$ does not depend on $\xi$ at all.  The same is true for the 
quarks, but now when $q (x) \rightarrow$ constant, that is when $\lambda_q = -1$, as 
seen in the $R_q^{ns} = 1$ result of Fig.~2(c).

All the scale dependence of the distributions is hidden in the $Q^2$ behaviour of the 
powers $\lambda (Q^2)$.  The position of the saddle point $N = \lambda$ in the 
Mellin integral (\ref{eq:a9}) moves to the right in the complex $N$ plane as $Q^2$ 
and so the off-diagonal \lq\lq enhancement\rq\rq~increases; in other words $R$ 
increases with $Q^2$.  A particular example is the double logarithm approximation 
when, in the singlet sector, the saddle point
\be
\label{eq:a24}
N \; = \; \lambda_g (Q^2) \; \simeq \; \sqrt{(\alpha_S/\pi) \: \ln (1/x) \: \ln 
(Q^2/Q_0^2)}.
\ee

In Fig.~2(d) we show the off-diagonal gluon distribution again, but now using a (more 
detailed) linear scale and comparing with the diagonal distribution $H (\bar{x}, 0)$ 
taken at fixed $\bar{x} = 2 \xi$, so as to avoid the extra $x$ dependence coming from 
the diagonal gluon in the denominator of the $R_g$ ratio.  This demonstrates that the 
extra $x$ dependence is responsible for the slight decrease observed in $R_g$ of 
Fig.~2(b) as $x \rightarrow 0$, and that the decrease is not due to the behaviour of 
$H_g (x, \xi)$.

The behaviour of the ratios at $x = \xi$ are explicitly
$$
R \; = \; \frac{H (\xi, \xi)}{ H (2 \xi, 0)}\; =\; \frac{2^{2\lambda+3}}{\sqrt{\pi}}
\frac{\Gamma(\lambda + 5/2)}{\Gamma(\lambda + 3 + p)}\, ,
$$
where $p = 0$ for quarks and $p = 1$ for gluons.  The ratios are plotted in Fig.~3 
 as a function of $\lambda$.  The vertical 
arrows shown on the plot indicate the values of $\lambda_g$ and $\lambda_q$ 
obtained from the gluon and sea quark distributions at $Q^2 = 4$ and 100~GeV$^2$ 
of a recent global (diagonal) parton analysis \cite{MRST}.  The plot can be used to 
find the enhancement of the cross section for the high energy diffractive 
electroproduction of vector mesons arising from off-diagonal parton effects.  The 
enhancement is given by $R_g^2$ where $R_g$ is the value of the gluon ratio at $x = 
\xi$, which is shown in Fig.~3, at the appropriate scale, that is at the appropriate value 
of $\lambda_g (Q^2)$.  For instance, for the photoproduction of $J/\psi$ and 
$\Upsilon$ at HERA the enhancement is about (1.15)$^2$ and (1.32)$^2$ 
respectively\footnote{In practice the diagonal distributions have more complicated 
forms than that assumed in (\ref{eq:a20}).  For instance if we were to input $xg \sim 
x^{- \lambda_g} (1 - x)^6$ in (\ref{eq:a19}) and to perform the $x^\prime$ 
integration numerically then we find $R_g$ increases from 1.32 to 1.41 for 
$\Upsilon$ photoproduction at HERA where $x \simeq 0.01$; the change in $R_g$ 
occurs because the $x$ sampled by the HERA data is not sufficiently small.  $R_g^2 
\simeq 2$ is in agreement with the previous estimates of the enhancement due to 
off-diagonal effects \cite{MR,MRT}.}, if we use a scale $M_V^2/4$,
 where $M_V$ is 
the mass of the vector meson.

From Figs.~2 and 3 we see that the off-diagonal or \lq\lq skewed\rq\rq~effect (the 
ratio $R$) is much stronger for singlet quarks than for gluons.  The explanation is 
straightforward.  At low $x$ the distributions are driven by the double leading 
logarithmic evolution of the gluon distribution.  At each step of the evolution the 
momentum fractions $x_i$ are strongly ordered ($x_1^\prime \gg x_1, x_2^\prime \gg 
x_2$ on Fig.~1).  For gluons it is just the \lq\lq last splitting function\rq\rq \ $P_{gg} 
(x_2, x_2^\prime; \xi)$ which generates the main $\xi$ dependence, or skewedness, of 
the distribution.  However for the sea or singlet quarks it is necessary to produce a 
quark with the help of $P_{qg}$ at the last splitting.  The splitting function $P_{qg}$ 
has no logarithmic $1/z = x_2^\prime/x_2$ singularity and so $x_2$ is the order of 
$x_2^\prime$.  Consequently both the splitting functions $P_{qg} (x_2, x_2^\prime; 
\xi)$ and $P_{gg} (x_2^\prime, x_2^{\prime\prime}; \xi)$ generate the asymmetry of 
the off-diagonal distribution.  Hence, at low $x$, the singlet quark has a much 
stronger off-diagonal effect than the gluon. \\

\section{Discussion}

In order to conclude that the conformal moments allow us to use the diagonal partons 
to uniquely determine the off-diagonal partons at small $x$ and $\xi$, including also 
their normalisation and $Q^2$ behaviour, it is necessary to consider some further 
points.

First we could worry that in the analytical continuation in $N$ of the conformal 
moments,
\be
\label{eq:a25}
O_N \; = \; \sum_k \: \xi^{2k} \: O_{Nk} \quad\quad {\rm with} \quad\quad 2k \: < \: 
N + 1,
\ee
 the higher ($k \geq 1$) terms will generate a singularity at $N > \lambda + 2k$.  
In such a case the small $x, \xi$ contribution would be driven by this singularity.  
However we show that such a singularity to the right of $N = \lambda + 2k$ cannot 
occur.  From the structure of the polynomials $R_N (x_1, x_2)$ of (\ref{eq:a3}) 
 and (\ref{eq:a5}), it is clear that there 
are no such singularities for integer $N > 2k$.  On the other hand a singularity at 
 non-integer $N = \beta + 2k$ would generate a function $f (x^\prime)$ of (\ref{eq:a9}) 
which depends on the ratio $\xi^{2k}/x^{\prime \beta + 2k}$.  After the convolution 
(\ref{eq:a10}) we would obtain a distribution which violates the polynomial condition 
\cite{J3},
\be
\label{eq:a26}
\int \: dx \: x^N \: H (x, \xi) \; = \; \sum_{k = 0}^{[(N + 1)/2]} \: A_k \xi^{2k},
\ee
which comes from Lorentz invariance (and the tensor structure of the operators).  
Thus the higher $\xi^2$ terms (with $k \geq 1$) in (\ref{eq:a13}) should die out with 
decreasing $\xi$.

A second consideration is that, from a formal point of view, we may add to the 
off-diagonal distribution any function
\be
\label{eq:a27}
\Delta H (x, \xi) \; = \; g (x, \xi) \: \theta (\xi - | x | )
\ee
which exists only in the time-like ERBL region~\cite{ERBL} with $| x | < \xi$.  Such a contribution 
$\Delta H$ remains in the ERBL region during evolution.  However $\Delta H$ 
disappears as $\xi \rightarrow 0$ and so it cannot be restored purely from diagonal 
partons.  A physical way to model such an ERBL contribution is to consider $t$ 
channel meson ($M$) exchanges of Fig.~4.  The contribution $\Delta H$ is then given 
by the leading twist wave function $\psi_M$ of the meson multiplied by the 
corresponding Regge exchange amplitude
\be
\label{eq:a28}
\Delta H^{\rm Reggeon} \; = \; \psi_M (x/\xi, Q^2) \xi^{- \alpha_M (t)} \: V (\mu^2 = 
Q_0^2; \xi; t).
\ee
The appropriate exchange is the $f_2$ meson which, in the constituent quark model, 
is formed from a $P$-wave $q\bar{q}$ state with $J^{PC} = 2^{++}$.  The Regge 
factor $\xi^{- \alpha_M}$ is the analogue of the $x^{- \eta}$ (or $\xi^{- \eta}$) factor 
in the non-singlet quark distribution $H^{ns} \sim x^{- \eta}$; in our notation of 
(\ref{eq:a20}) and Fig.~2 with $xq^{ns} \sim x^{- \lambda_{ns}}$ we have 
$\lambda_{ns} = \eta - 1$.  Phenomenologically we expect that $\eta \sim \alpha_M 
(0) \sim 0.5$.  The key factor in (\ref{eq:a28}) is $V$ which specifies the coupling of 
the Reggeon to the proton.  From Regge phenomenology the vertex factor $V$ was 
extracted for the diagonal case where the ERBL domain does not exist.  Let us try to 
estimate a possible ERBL contribution to the off-diagonal distributions.  The value of 
the pion-nucleon $\Sigma$-term at low scales determines the number of current quarks 
and antiquarks in the nucleon to be \cite{QQ}
\be
\label{eq:a29}
\langle N | \bar{q}q | N \rangle \; \simeq \; 8.
\ee
Allowing for valence quarks, this implies that the average number of $q\bar{q}$ pairs 
is about 2.5.  At such low scales the partons are distributed more or less uniformly in 
the whole $(- 1, 1)$ interval and so the probability to find two partons in the ERBL 
domain $(- \xi, \xi)$ is of the order of $\xi^2$.  Such a $\Delta H$ is a negligible $O 
(\xi^2)$ contribution at small $\xi$ in agreement with our decomposition of the 
conformal moments.

So far our distributions enable us to calculate the imaginary part of the amplitude, say 
for Compton scattering\footnote{To be specific we consider the case with $t \leq 0$, 
$q^2 \leq 0$ and $q^{\prime 2} \leq 0$.} with incoming and outgoing photon 
virtualities $q^2 = - Q^2$ and $q^{\prime 2} = - Q^{\prime 2}$.  At small $x$ and 
$\xi$ it turns out that the real part of the amplitude may be calculated easily using a 
dispersion relation in the centre-of-mass energy squared $W_s^2 = (p + q)^2$.  Let us 
consider the cut in the right-half $W_s$ plane, that is the discontinuity for $W_s^2 > 
0$.  For fixed $t, Q^2$ and $Q^{\prime 2}$, the ratio $r = x/\xi$ is fixed as well, since 
$(x + \xi)/(x - \xi) = Q^2/Q^{\prime 2}$.  Thus the energy squared may be written
\be
\label{eq:a30}
W_s^2 \; = \; (1 - x) \: \frac{Q^2}{x + \xi} \; = \; (1 - r \xi) \: \frac{Q^2}{1 + r} \: 
\frac{1}{\xi}.
\ee
However we must take into account the cuts in both right and left half-planes, that is 
the $s$ and $u$ channel cuts.  The left-hand cut corresponds to the $u$ channel 
process (obtained by the interchange $p \leftrightarrow - p^\prime$) with energy squared
\be
\label{eq:a31}
W_u^2 \; = \; - (1 + x) \: \frac{Q^2}{x + \xi}.
\ee
The unpolarized deeply virtual Compton amplitude is the sum of the $s$- and $u$-
channel terms, $A = A_s + A_u$, and appears to have even signature, that is $A$ 
 is crossing 
symmetric.  Strictly speaking at large $x$ and $\xi$ there is some asymmetry (since 
$W_u^2 \neq - W_s^2$), which may be considered as the odd signature contribution 
and should be treated appropriately in the dispersion integral.  However the situation 
is particularly simple at small  $x \ll 1$, where $(1 \pm x) \simeq 1$.  Then we 
may write the whole amplitude $A \propto (W^2)^\lambda$, with the help of the even 
signature factor
\be
\label{eq:a32}
{\cal S}^+ \; = \; \frac{1}{2} (1 + (-1)^\lambda),
\ee
in the form
\be
\label{eq:a33}
A \; = \; \frac{i}{2} \: {\rm Im} A (1 + e^{- i \pi \lambda}).
\ee
Moreover for small $\lambda$ we have
\be
\label{eq:a34}
{\rm Re} A \; \simeq \; \frac{\pi \lambda}{2} \: {\rm Im} A.
\ee

Strictly speaking the conformal moments $O_N$ only renormalize multiplicatively, as 
in (\ref{eq:a8}), at leading order (LO).  Due to a conformal anomaly at next-to-leading 
(NLO) the moment $O_N$ mixes, on evolution, with moments $O_{N^\prime}$ with 
$N^\prime < N$ \cite{STAN}.  The mixing is taken into account by a matrix 
$\mbox{\boldmath $B$}_{NN^\prime}$, which obeys its own evolution equation 
\cite{BM}.  Of course the mixing is absent in the diagonal case when $\xi \rightarrow 
0$, whereas for non-zero $\xi$ we have
\be
\label{eq:a35}
O_N^{\rm NLO} \; = \; \sum_{N^\prime = 0}^N \: \mbox{\boldmath 
$B$}_{NN^\prime} \: O_{N^\prime}^{\rm NLO (diag)}  \: \xi^{N 
- N^\prime}
\ee
where $O_N, O_{N^\prime}$ and $\mbox{\boldmath $B$}_{NN^\prime}$ all depend on $\alpha_S 
(Q^2)$.  Thus in the small $\xi \ll 1$ limit we can safely use expressions 
(\ref{eq:a15}) and (\ref{eq:a19}) for $H (x, \xi)$ even at NLO.

In summary, in the low $\xi$ region we can use expressions 
 (\ref{eq:a15}) and (\ref{eq:a19}) to reliably predict the off-diagonal 
distributions $H (x, \xi)$ in terms of the diagonal partons 
 at any scale.  All that is required is a two-fold 
integration.  The expected accuracy is of the order of $\xi^2$.  As a specific example 
we assumed in (\ref{eq:a20}) that the diagonal partons had a power-like $x^{-
\lambda}$ behaviour for small $x$.  In this case one integration can be done 
analytically and we have even simpler expressions for $H_q$ and $H_g$, see 
(\ref{eq:a21}).  The results are shown in Figs.~2 and 3 and allow the off-diagonal 
distributions to be determined for any small $x, \xi$ values at any scale.  One 
important consequence is that data for processes, which are described by off-diagonal 
distributions, can be included in a global analysis to better constrain the low $x$ 
behaviour of the (conventional) diagonal partons. \\

\vspace*{1cm}
\noindent {\large \bf Acknowledgements}

We thank Lev Lipatov for helpful discussions.  This work was supported in part by the 
Royal Society, INTAS (95-311), the Russian Fund for Fundamental Research (98 02 
17629),  the Polish State    
Committee for Scientific Research grant No.~2~P03B~089~13
and by the EU Fourth Framework Programme TMR, Network \lq QCD and 
the Deep Structure of Elementary Particles\rq~contract FMRX-CT98-0194 (DG12-MIKT)
and (DG12-MIHT).

\newpage  
\begin{figure}
   \vspace*{-1cm} 
    \centerline{ 
     \epsfig{figure=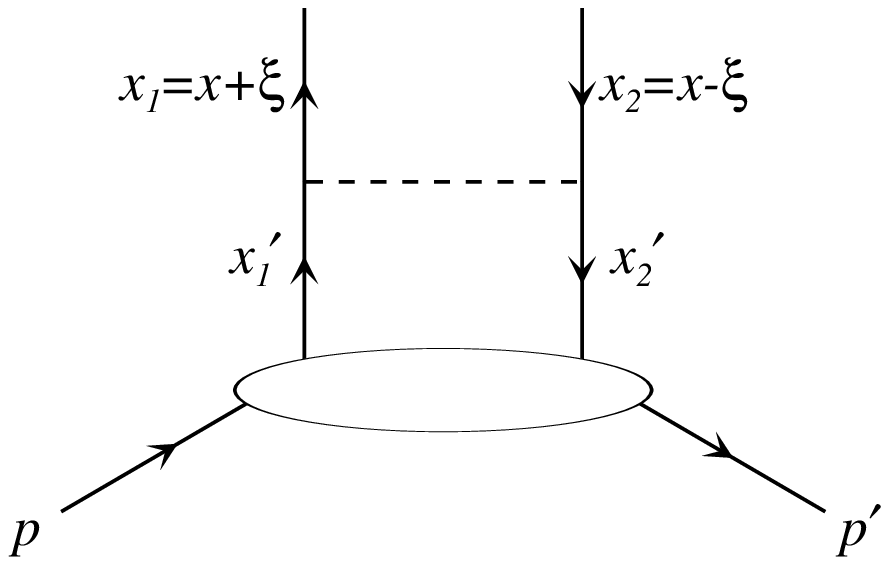,width=15cm} 
               } 
    \vspace*{-0.5cm} 
\caption{A schematic diagram showing the variables for the off-diagonal parton 
distribution $H (x, \xi)$ where $x_{1,2} = x \pm \xi$.} 
\end{figure}

\newpage  
\begin{figure}
   \vspace*{-1cm} 
    \centerline{ 
     \epsfig{figure=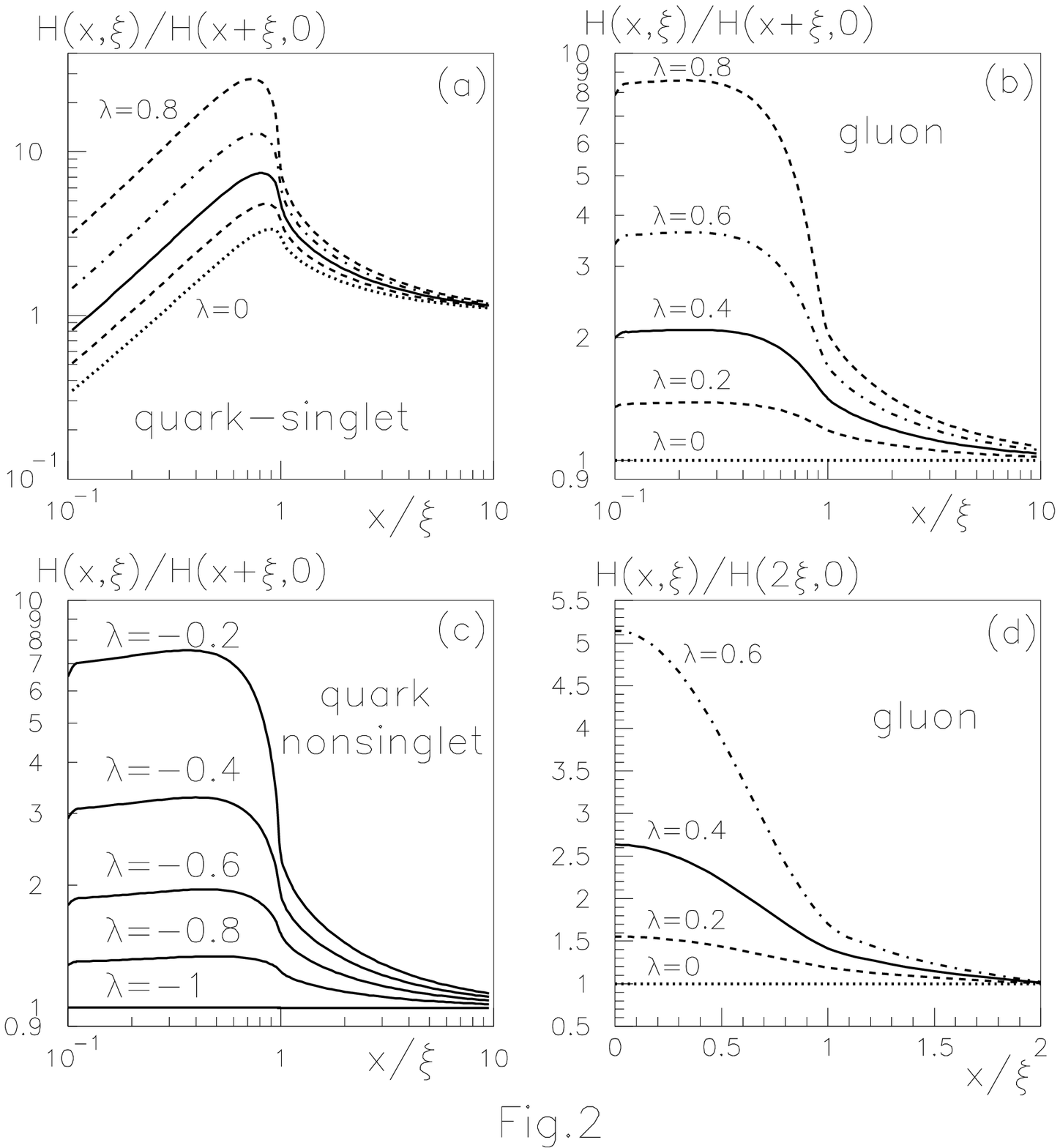,width=15cm} 
               } 
    \vspace*{-0.5cm} 
\caption{Predictions at small $x$ and $\xi$ for the ratio of off-diagonal ($H (x, 
\xi)$) to diagonal $(H_q (\bar{x}, 0) = f_q (\bar{x}), H_g (\bar{x}, 0) = \bar{x} f_g 
(\bar{x}))$ parton distributions.  The diagonal partons are taken to have the form $x f 
(x) = N x^{- \lambda}$.  Plots (a), (b) and (c) show the quark singlet, gluon and quark 
non-singlet ratios taking $\bar{x} = x + \xi$ as the argument of the diagonal partons.  
Plot (d) shows the gluon ratio again but versus a linear scale and with argument 
$\bar{x} = 2 \xi$ so as to display the $x$ behaviour of $H_g (x, \xi)$.} 
\end{figure}

\newpage  
\begin{figure}
   \vspace*{-1cm} 
    \centerline{ 
     \epsfig{figure=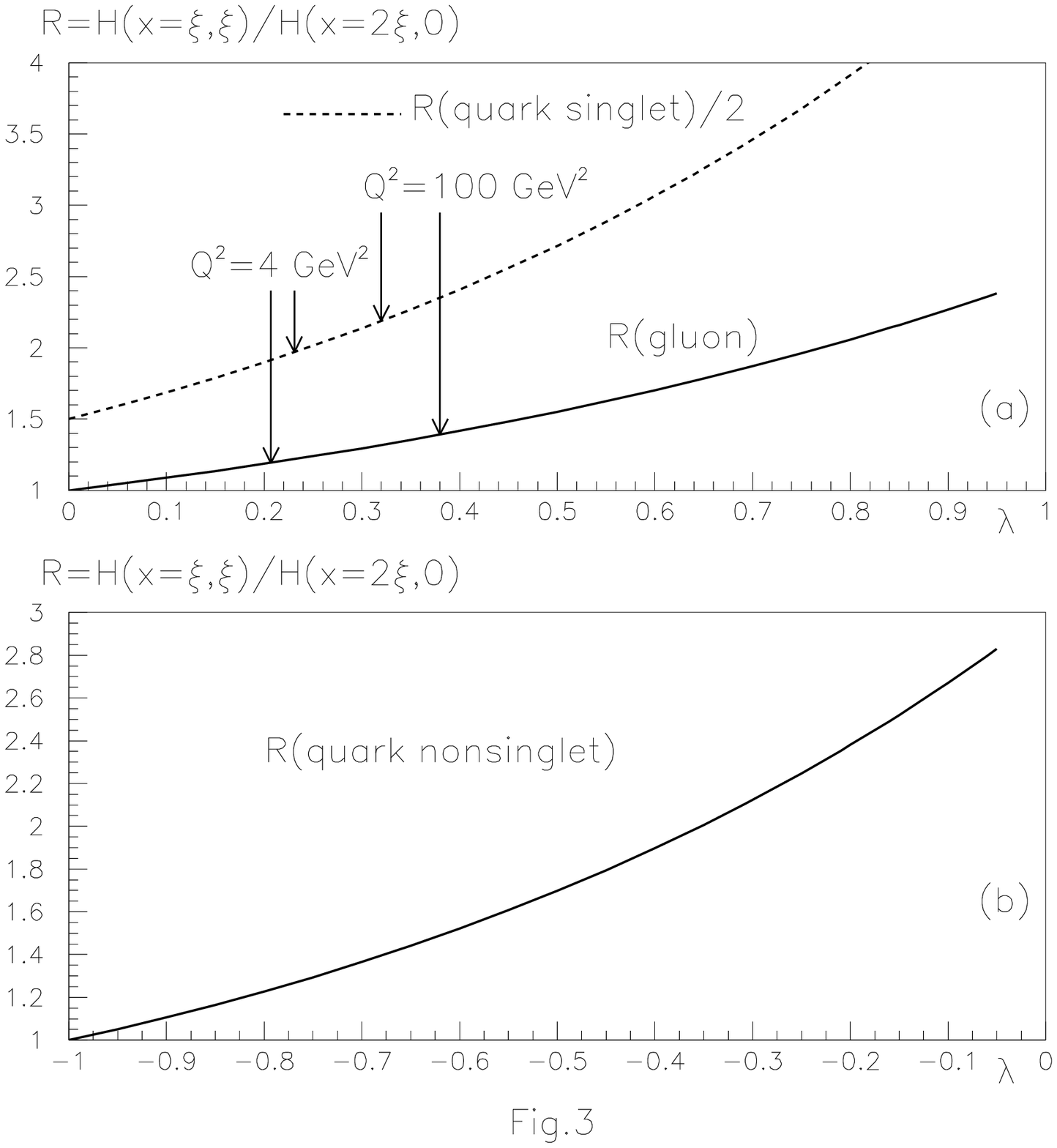,width=15cm} 
               } 
    \vspace*{-0.5cm} 
\caption{The off-diagonal to diagonal ratio, $R$, at small $x$ and $\xi$ versus 
the power $\lambda$ which specifies the $x^{- \lambda}$ behaviour of the input 
diagonal parton as in (\ref{eq:a20}).  Note that the quark singlet ratio has been divided 
by 2.  The vertical arrows indicate the values of $\lambda$ found in a global parton 
analysis \cite{MRST} at $Q^2 = 4$ and 100~GeV$^2$.} 
\end{figure}

\newpage  
\begin{figure}
   \vspace*{-1cm} 
    \centerline{ 
     \epsfig{figure=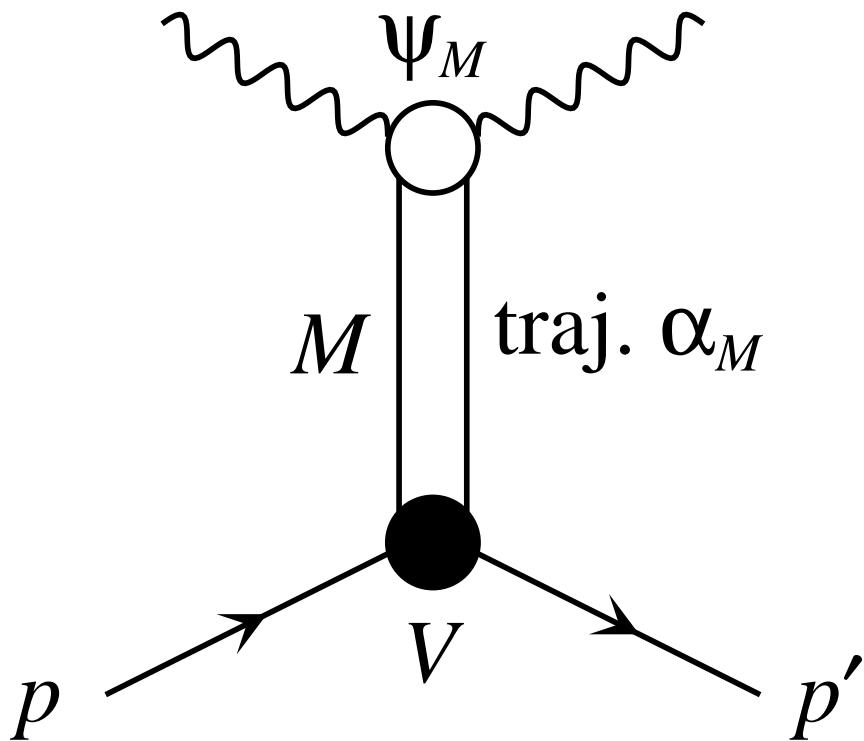,width=15cm} 
               } 
    \vspace*{-0.5cm} 
\caption{Meson ($M$) Regge exchange indicating the structure of the 
off-diagonal contribution of (\ref{eq:a28}).} 
\end{figure}

\end{document}